\documentclass[aoas,MSNbibl,nameyear,dvips]{arximspdf}
\usepackage{mathbh}
\usepackage{dcolumn}
\usepackage{graphicx}

\doi{10.1214/11-AOAS521}
\volume{6}
\issue{2}
\pubyear{2012}
\firstpage{772}
\lastpage{794}

\makeatletter

\setattribute{copyright}{owner}{In the Public Domain}
\newcolumntype{d}[1]{D{.}{.}{#1}}
\newcolumntype{e}[1]{D{.}{}{#1}}
\newcommand{\mathbbm}{\mathbh}
\renewcommand{\epsilon}{\varepsilon}
\makeatother

\begin{document}
\begin{frontmatter}

\title{Analyzing establishment nonresponse using an~interpretable
regression tree model with linked administrative data}
\runtitle{Analyzing establishment nonresponse}

\begin{aug}
\author[A]{\fnms{Polly} \snm{Phipps}\ead[label=e1]{toth.daniell@bls.gov}}
\and
\author[A]{\fnms{Daniell} \snm{Toth}\corref{}\ead[label=e2]{phipps.polly@bls.gov}}
\runauthor{P. Phipps and D. Toth}
\affiliation{Bureau of Labor Statistics}
\address[A]{Office of Survey Methods Research\\
Bureau of Labor Statistics\\ Suite 1950\\
Washington, DC 20212\\USA\\
\printead{e1}\\
\hphantom{\textsc{E-mail: }}\printead*{e2}} 
\end{aug}

\received{\smonth{11} \syear{2010}}
\revised{\smonth{9} \syear{2011}}

%
\begin{abstract}
To gain insight into how characteristics of an establishment are
associated with nonresponse,
a recursive partitioning algorithm is applied to the Occupational
Employment Statistics May 2006 survey data
to build a regression tree.
The tree models an establishment's propensity to respond to the survey
given certain establishment characteristics.
It provides mutually exclusive cells based on the characteristics with
homogeneous response propensities.
This makes it easy to identify interpretable associations between the
characteristic variables
and an establishment's propensity to respond,
something not easily done using a logistic regression propensity model.
We test the model
obtained using the May data against data from the November 2006
Occupational Employment Statistics survey.
Testing the model on a disjoint set of establishment data with a very
large sample size $(n=179\mbox{,}360)$
offers evidence that the regression tree model accurately describes
the association between the establishment characteristics and the
response propensity for the OES survey.
The accuracy of this modeling approach is compared to that of logistic
regression through simulation.
This representation is then used along with frame-level administrative
wage data linked to sample data
to investigate the possibility of nonresponse bias.
We show that without proper adjustments the nonresponse does pose a
risk of bias and is possibly nonignorable.
\end{abstract}

%

\end{frontmatter}

\section{Introduction}

Survey nonresponse and associated risks of nonresponse bias are major
concerns for government agencies
and other organizations conducting the establishment surveys that
produce a nation's economic statistics.
Survey methodologists, as well as survey programs and sponsors,
consider response rates
an important measure of data quality.
While establishment surveys have not shown a consistent downward trend
in response rates,
achieving and maintaining a high response rate has become more
difficult over time
[\citet{Petetal}].
Increased efforts on the part of agencies and organizations have
stemmed response rate declines in many cases.
For example, most Bureau of Labor Statistics (BLS) establishment
surveys have
dedicated resources to maintain their response rates through either
design changes or
increased collection efforts over the past decade.
Determining exactly where to focus these efforts is a subject of this paper.

Lower response rates also may be associated
with nonresponse bias, if respondents and nonrespondents differ on
survey outcomes.
Further, adjusting for large amounts of nonresponse may induce more
variance in the estimator
[\citet{LitVar05}], as well as a loss of confidence in the
data by the stake-holders.
Investigation of differences in characteristics of respondents and
nonrespondents is an important
survey data quality procedure.
These differences in characteristics can be used to decide where to
direct additional resources and efforts
in the data collection process and in adjusting the estimates after the
data is collected.
However, there are few studies in the literature that examine the
association between
establishment characteristics and survey response.
In addition, analysis of whether respondent differences are
systematically related to
survey outcomes is critical to developing post-survey adjustments to
account for nonresponse bias.

The BLS Occupational Employment Statistics survey (OES) is a semi-annual
establishment survey measuring occupational employment and wage rates
for wage and
salary workers by industry for the U.S., states, certain U.S.
territories, and metropolitan statistical areas.
This voluntary survey of establishments with one or more employees is
primarily conducted by mail.
The OES attains one of the highest response rates of all BLS
establishment surveys, approximately 78 percent.
Even with a high response rate, \citet{PhiJon07} indicate that a
number of important variables
related to establishment characteristics are associated with the
likelihood to respond to the OES.
The authors find that establishment characteristics have a greater association
with the probability of OES response compared to characteristics of the
survey administration.
Characteristics of establishments thought likely to be associated with
survey response propensity
include the following: the size of the establishment, measured by the
number of employees;
the industry classification of the establishment; whether an
establishment is part of a larger firm;
and the population size of the metropolitan area in which the
establishment is located,
among others [\citet{aa}].
These variables are usually available on government and private
sampling frames.
Since many of these characteristics can also be associated with an
establishment's wages,
an important OES outcome variable, nonresponse bias is a potential concern.

The primary goal of this work is to identify a set of characteristics
that partition the establishments into groups of establishments with
similar response rates.
For example, we wish to accurately describe the propensity for a given
establishment to respond to the OES survey based on its class membership.
Classes are defined by certain characteristics known for all sampled
establishments.
This would allow for the easy identification of establishments that are
more likely to be OES nonrespondents and may warrant additional
collection effort.

To analyze OES survey response, we use the following inferential framework.
Suppose a sample is drawn from a finite population $\{(Y_1, \mathbf
{X}_1), \ldots,\break (Y_N, \mathbf{X}_N)\}$ indexed by the set $U.$
Let $S \subset U$ be the set of sampled elements.
For each $i \in U,$ let $R_i = 1$ if unit $i$ responds to the OES
survey, if selected in the sample,
and zero otherwise.
Note that the value of $R_i$ is only known for those $i$ selected in
the sample.
We assume that each $R_i$ is an independent Bernoulli random variable with
%
\begin{equation}\label{Bernoulli}
P (R_i =1 | \mathbf{x}=\mathbf{x}_i ) = p(\mathbf{x}_i).
\end{equation}
The function $p(\mathbf{x}_i)$ is called the response propensity of
unit $i.$
Since $R_i$ is a Bernoulli random variable, the response propensity,
equation (\ref{Bernoulli}), can be modeled by estimating the
conditional mean
using the equality $p(\mathbf{x}_i)=E(R_i|\mathbf{x}_i).$

A common tool used to model the response propensity is the parametric
logistic regression model
[see \citet{LitVar05}; \citet{Lit86}; \citet{RosRub83}].
The response propensity for unit $i$, given characteristic variables
$\mathbf{x}_i,$ is modeled by
%
\begin{equation}\label{logistic}
p(\mathbf{x}_i)= (1 + \exp\{-z_i\} )^{-1},
\end{equation}
where $z_i = \beta\mathbf{x}_i = \beta_0 + \beta_1 x_{i1} +\cdots+
\beta_p x_{ip}.$
Mutually exclusive response cells are defined for which the propensity
is approximately equal according to the
modeled quantiles; these cells are used in adjusting estimates for
nonresponse bias [\citet{VarLit}].
This is the case when either weighting or calibration is used to adjust
for nonresponse
[see \citet{Lit82}; \citet{KotCha10}].

The identification of interpretable response cells is often challenging
using this model based method
[\citet{EltYan97}; \citet{KimKim07}].
Establishments in the same response cell often have very different
establishment characteristics.
This becomes a major difficulty when variables are
continuous and their association with the response rate is not
monotonic or includes interaction effects.
In the case of the OES, the response rate will be shown to be
especially low for establishments in
a small number of industries, but this difference depends on the size
of the establishment,
among other characteristics.
The response model produced from the logistic regression method using
OES data included
many significant interaction effects; see equation (\ref{L1}).

In addition to being difficult to interpret, the logistic regression
model may encounter problems with adequacy of
fit for the specified model,
%
\begin{equation}\label{linear}
\operatorname{logit}(\mathbf{x}) = \beta\mathbf{x}.\vadjust{\goodbreak}
\end{equation}
For example, the specified vector $\mathbf{x}$ may fail to include
predictors that fully account for
curvature or interactions that may be important for some of the types
of nonresponse, and thus suffer from lack of fit (see Figure~\ref{smooth}).

In contrast, regression trees are a nonparametric approach that results in
mutually exclusive response cells, $C_1, \ldots, C_{k+1},$ based on
similar establishment characteristics
containing units with homogeneous propensity scores.
To estimate $p(\mathbf{x})$, the regression tree estimates the mean
value $p(\mathbf{x})$ for each category, $C_j,$ separately, by
%
\begin{equation}
\biggl(\sum_{i\in S}{\mathbbm{1}_{\{\mathbf{x}_i \in C_j\}}} \biggr)^{-1} \sum
_{i\in S}{ R_i\mathbbm{1}_{\{\mathbf{x}_i \in C_j\}}}.
\end{equation}

See \citet{SchdeN} and \citet{GokJudMos92}
for examples of the use of recursive partitioning algorithms for
producing response cells.
In this way, previous works use regression trees as a substitute for
logistic regression to conduct nonresponse adjustment.
Here, we use the interpretability of the regression tree structure to
examine the association between establishment
characteristics and survey response.

The resulting tree model is easily cast as a linear regression of the form
%
\begin{equation}\label{simp_eq}
p(x)= \beta_0 + \beta_1 S_1(x) +\cdots+ \beta_k S_k(x),
\end{equation}
where $S_i$ for $i=1,\ldots, k$ are the indicator functions of whether
the establishment
has the defined characteristic or not [see \citet{TotEltN2} or
\citet{LeBTib98}].
These indicator functions define the splits in formation of the trees.
In this form, the coefficients are easy to interpret as the association
between a specific characteristic
with the establishment's propensity to respond to the OES survey.

It is clear that the resulting tree model partitions the establishments
into one of $k+1$ classes defined by which splits the establishment's
characteristics satisfy.
The response propensity for a given establishment is then simply the
base propensity $\beta_0,$
plus the sum of all the coefficients $\beta_i$ for which the establishment's
characteristics satisfy split $S_i$.
Equation (\ref{simp_eq}) can be written as
%
\begin{equation}\label{estimator}
p(\mathbf{x})= \mu_1 C_1(\mathbf{x}) +\cdots+ \mu_{k+1}
C_{k+1}(\mathbf{x}),
\end{equation}
where $C_i$ is the indicator function of whether a given
establishment's characteristics designate
membership to class $i.$
For example, if class $C_i$ is defined as satisfying the first $j$
splits and not satisfying the rest, then
the estimated response propensity of establishments in that class is
given by
\[
\mu_i= \beta_0 +\cdots+ \beta_j.
\]
This form allows us to easily define nonresponse adjustment groups
based on known establishment characteristics and
identify groups of establishments that may require extra collection
effort for the OES survey.

To build the nonparametric regression tree model, we use a recursive
partitioning algorithm
which minimizes the estimated squared error of the estimator defined by
equation (\ref{estimator})
[see Gordon and Olshen (\citeyear{GorOls78}, \citeyear{GorOls80})].
The splits, and therefore the variables for our model, are chosen using
a~method based on cross-validation estimates of the variance.
A test of the model obtained using the May 2006 sample is performed by
estimating the response rates
[the parameters in equations (\ref{simp_eq}) and (\ref{estimator})] on
the November 2006 sample.
The OES samples are selected so that the set of establishments in the
November sample
represent a disjoint set of establishments from the May sample (the
data on which the model was obtained).
This procedure is more fully explained next in Section~\ref{recursive
partitioning}.

Section~\ref{analysis} describes the OES sample frame, survey, and data,
and analyzes the response patterns of establishments using regression trees.
Section~\ref{bias} explores possible nonresponse bias.
A discussion of the results is contained in Section~\ref{discussion}.
An evaluation of the performance of the nonparametric regression tree
relative to the parametric logistic regression
using several different response mechanisms is given in the \hyperref[sec:sim]{Appendix}.

\section{Description of the recursive partitioning algorithm}\label
{recursive partitioning}

A recursive partitioning algorithm is used to build a binary tree that
describes the association between an establishment's characteristic
variables and its
propensity to respond to the OES. A recursive partitioning algorithm begins
by splitting the entire sample, $\mathcal{S}$, into two subsets,
$\mathcal{S}_1$ and $\mathcal{S}_2$,
according to one of the characteristic variables.
For example, the partitioning algorithm could divide the sample of
establishments in the OES
into establishments that have more than 20 employees and those that
do not. The desired value (in this case the proportion of respondents)
is then estimated for each subset separately.
This procedure is repeated on each subset (recursively) until the resulting
subsets obtain a predefined number of elements.
At each step, the split that results in the largest decrease in the
estimated mean squared error
for the estimator is chosen, from among all possible splits on the
auxiliary variables.
This is the same criteria used in the classification and regression
tree (CART) procedure explained in
\citet{Breetal84}.
This results in a tree model $p(x)$ of the forms (\ref{simp_eq}) and
(\ref{estimator}).

In a series of papers by  Gordon and Olshen (\citeyear{GorOls78}, \citeyear{GorOls80}) for the simple
random sample case, and \citet{TotEltN1} for the complex sample case,
asymptotic consistency was established for the mean estimator based on
a recursive partitioning algorithm.
The consistency proofs require that the resulting subsets all have at
least a minimum number of sample elements, and,
as sample size increases, the minimum size also increases.
The minimum size must increase at a rate faster than $\sqrt{n}.$
Thus, we required a minimum sample size of $n^{5/8}$ in each subset.

In order to obtain a more parsimonious model, we retained only the
first $k$ splits.\vadjust{\goodbreak}
We choose $k$ using a procedure based on the 10-fold cross-validation.
This is done by first dividing the sample $\mathcal{S}$ of size $n$
into 10 groups, $G_1, G_2,\ldots,G_{10},$ each of size
$n/10$ by simple random sampling.
For a given $k,$ estimate a regression tree model $p_j$ with $k$ splits
using the sample data, excluding the set $G_j.$
To estimate the mean squared prediction error of the tree model on the
set $G_j$, we compute
\[
e_j=10n^{-1}\sum_{i \in G_j}{ \{R_i-p_j(\mathbf{x}_i) \}^2}.
\]
For a tree with $k$ splits, the expectation of the overall mean squared
prediction error,
\[
E [\{R_i - p(\mathbf{x}_i)\}^2 ],
\]
is then estimated by
\[
\epsilon^2_k=10^{-1}\sum_{j=1}^{10}{e_{j}}.
\]
Let $\bar{R}=n^{-1}\sum_{i \in\mathcal{S}}R_i$ be the estimated
overall response rate.
Defining
\[
\alpha^2_0 =n^{-1}\sum_{i \in\mathcal{S}} (R_i - \bar{R} )^2,
\]
we estimate the relative mean prediction error of the tree model with
$k$ splits by $r^2_k = \epsilon^2_k / \alpha^2_0.$
The variance of $r_k$ is estimated by
\[
\sigma^2_k =(9\alpha^2_0)^{-1}\sum_{j=1}^{10}{ (e_j - \epsilon_k )^2}.
\]
Both $r_k$ and $\sigma_k$ are calculated for increasing values of $k$ until
the model with $k+1$ fails to reduce the estimated relative overall
prediction error, $r_{k+1},$ by more than one times its
estimated standard deviation $\sigma_{k+1}$,
\[
|r_{k+1} - r_{k}| \geq\sigma_{k+1}  \qquad \mbox{where } k = 0, 1, 2, \ldots.
\]
Note that this procedure represents a more conservative approach (fewer
splits) than the one advocated
by \citet{Breetal84} in the CART procedure,
which leads to a model with a larger number of splits.

We emphasize that the main objective is to identify and understand
the characteristics of an establishment that are most strongly
associated with the
propensity to respond to the OES survey and not to adjust the estimator
for nonresponse bias.
With this goal in mind, the following course of action seems reasonable.
First, adopt this more conservative approach to modeling over getting
the most accurate propensity prediction.
Second, build the regression tree model ignoring the sampling
design.

The conservative approach to modeling should help insure against overfitting.
That is, only features that\vadjust{\goodbreak}
are strongly associated with response are likely to be identified by
the model.
Using only characteristics with a relatively large association with the
response rate
leads to more stable estimators of those effects.
It also produces a smaller number of possible categories, making it
easier to explain which establishments
are likely to require additional nonresponse follow-up effort.
Likely one can refine the classification further by taking a more
aggressive modeling approach at the risk of obtaining a less stable model.

Whether to account for the sample design by using weights in the
modeling of nonresponse depends on the intended
use of the model and is the subject of ongoing research.
Ignoring the sampling design in this case makes sense when we consider
that our population of inference is not the
sampled population but future samples of the OES selected with this
same design.

We would like to point out that this procedure could be adopted for a~repeated survey like the OES,
or for nonrepeated surveys or surveys in which the sample design changes.
In these situations, the population of inference is the target
population and not just the sampled ones, and the design is relevant.
Incorporating the sample design information in this method can be done by
building a consistent regression tree estimator using a weighted
estimator described in \citet{TotEltN1}.
This estimator is proven to be consistent, assuming the sample design
satisfies certain conditions.
The cross-validation can then be done using the weighted
cross-validation procedure proposed by \citet{OpsMil05}.

Recursive partitioning algorithms represent a nonparametric approach to
modeling a relationship between a response variable and a set of
characteristic variables.
However, to test the accuracy of the model,
we use the parametric forms given by equations (\ref{simp_eq}) and (\ref
{estimator})
of the resulting regression model.
This is done by re-estimating the linear coefficients using the
November OES sample, where all the establishments
in this new sample are disjoint from the May OES sample used to build
the model.

This test was possible because establishments for the OES are selected
in panels of separate establishments.
That is, the set of establishments in the November 2006 OES sample is
not in the May 2006 OES sample.
Therefore, the data used to test the model are completely disjoint from
the data from which the splits were obtained.
By comparing the May and November estimated coefficients of equation
(\ref{simp_eq}), we can assess how well
the model quantifies the association of each split with the
establishment's response propensity.
In addition, comparing the coefficients of equation (\ref{estimator})
for May and November,
we can check the accuracy of the model in predicting the
establishment's response propensity.

\section{Analysis of nonresponse in the OES using regression
trees}\label{analysis}

The semi-annual OES survey is conducted by state employment
workforce agencies in cooperation with the BLS.
For survey administration purposes, the state OES offices are grouped
into six regions.
Each region has a BLS office, and BLS personnel guide, monitor, and
assist the state offices.
The OES is primarily a mail survey; the initial mailing is done by a
central mail
facility, with three follow-up mailings sent to nonrespondents.
Additionally, state OES offices follow up with nonrespondents by telephone.
In 2006, approximately 72\% of establishments responding to the survey
provided data by mail,
12\% by telephone, 7\% by email, 4\% by fax, and the remainder provided
data in other electronic forms.
Regional office personnel directly collect a small proportion 
of OES data through special arrangements maintained with
multi-establishment firms
(referred to as central collection).
These establishments represented 8\% of total employment in May 2006.
Firms using this arrangement usually provide data for their sampled
establishments in an electronic format.

The survey is conducted over a rolling 6-panel semi-annual (or 3-year) cycle.
Each panel's sample contains approximately 200,000 selected establishments.
Over the course of a 6-panel cycle, approximately 1.2 million
establishments are sampled.
The sample is drawn from a universe of about 6.5 million establishments
across all nonfarm industries
and is stratified by geography, industry, and employment size.
The sample frame comes from administrative records: quarterly state
unemployment insurance (UI) tax reports
filed by almost all establishments.

The data used for the recursive partitioning algorithm are the May 2006
OES semi-annual
sample of 187,115 establishments
in the 50 states and District of Columbia.\footnote{We exclude federal
government establishments,
as the data are not collected at the
establishment level: one data file is provided to BLS by the Office of
Personnel Management.
State government establishments are collected in the November survey
panel and not included in this study.}
The data include variables measuring establishment characteristics for
all sample members,
including those that did not respond to the survey.
These variables, described in detail in Table~\ref{tab1}, are as follows:
employment size ($\mathit{EMPL}$),
industry supersector ($\mathit{IND}$),
metropolitan statistical area population ($\mathit{MSA}$), age of the
establishment in years ($\mathit{AGE}$),
the number of establishments with the same national employer
identification number ($\mathit{MULTI}$),
and whether the establishment provides support services to other
establishments within a firm ($\mathit{AUX}$).
All of these variables exist or were constructed from the BLS Quarterly
Census of Employment (QCEW)
establishment frame, which derives its data from the quarterly state UI
administrative tax records.

%
\begin{table}
\caption{Variables used along with type and values}\label{var_table}
\label{tab1}\vspace*{-3pt}
\begin{tabular}{@{}ll@{}}
\hline
\textbf{Variable name}&\textbf{Value}\\
\hline
$\mathit{EMPL}$ & Integer number of employees \\
$\mathit{IND}$ & 11 supersector categories following NAICS \\
& (1) natural resources and mining, (2) construction, (3)
manufacturing \\
& (4) trade/transportation/utilities, (5) information, (6) finance \\
& (7) professional and business services, (8) education and health \\
& (9) leisure and hospitality, (10) other services, (11) local
government\\
$\mathit{MSA}$ & 6 categories based on area population size\\

& (1) non-MSA, (2) 50-149,999, (3) 150-249,999 , (4) 250-499,999\\
& (5) 500-999,999, (6) 1,000,000$+$\\
$\mathit{AGE}$ & Real number of age in years calculated from the first
Unemployment\\
& Insurance liability date \\
$\mathit{MULTI}$ & Integer number of multi-establishments with same
national employer \\
& identification number\\
$\mathit{AUX}$ & Indicator whether establishment provides support
services to other \\
& establishments in the firm\\

$\mathit{MANDATORY}$ & Indicator whether the establishment is located
in one of three states\\
& that makes completion of the survey mandatory by law: OK, NC, SC\\
$\mathit{REGION}$& 6 categories \\
& denotes one of six BLS regional offices that assist the state office
\\
& responsible for collecting the establishment's data \\
$\mathit{CC}$ & Indicator whether a regional office attempted to
collect the data\\
&directly\\
\hline
\end{tabular}
\vspace*{-3pt}
\end{table}

In addition, Table~\ref{tab1} includes three variables of interest available in
the data for each establishment,
$\mathit{MANDATORY}$, $\mathit{REGION}$, and
$\mathit{CC}$, that are characteristics of the survey administration, not the
establishment.
The variable $\mathit{MANDATORY}$ indicates whether the establishment is located
in one of three\vadjust{\goodbreak} states that make
completion of the survey mandatory by law.
The variable $\mathit{REGION}$ denotes one of six BLS regional offices
that assist the state office that is responsible for collecting the
establishment's data.
The variable $\mathit{CC}$ indicates whether the data were centrally collected
or not,
that is, whether the data collection method was through a regional
office attempting to collect the data directly
through their special arrangements with some multi-establishment firms,
or whether a mailed survey form
was used (most establishments).

We first performed the recursive partitioning on the data with all the
described variables.
The estimated response rates and establishment characteristics
associated with response rates
depended on whether or not the data were centrally collected.
This is not surprising, as these firms have made the effort to request
that the BLS contact
a single representative for all establishments in the firm that are
selected into the OES sample.
The BLS regional office then coordinates the data collection for these
firms, and they are
considered a select group of firms.

Because of this interaction, and since our interest is to classify
characteristics of an establishment that
are strongly associated with the propensity to respond to a specific
method of data collection,
subsequent analyses were carried out separately for the survey- and
centrally collected establishments.\vadjust{\goodbreak}

\subsection{Survey-collected establishments}

Of the 187,115 establishments in the sample, the vast majority,
179,000, were not centrally collected.
These establishments were mailed a survey to be completed and returned
with the requested data.
In order to identify sets of establishments
with homogeneous response propensities for the survey based on the
establishment characteristics,
we recursively partitioned the data using the algorithm described above
on the characteristic variables:
$\mathit{EMPL}$, $\mathit{IND}$, $\mathit{AGE}$, $\mathit{MULTI}$, $\mathit{AUX}$, $\mathit{MSA}$, $\mathit{REGION}$, and $\mathit{MANDATORY}$.

%
\begin{table}
\tabcolsep=0pt
\tablewidth=160pt
\caption{Relative estimated error by number of splits,
survey-collected establishments, May 2006}\label{var_table_tree1}
\label{tab2}\begin{tabular*}{160pt}{@{\extracolsep{\fill}}lcc@{}}
\hline
\textbf{Split}&\textbf{Estimate}&\textbf{Standard error}\\
\hline
\hphantom{0}0 & 1.0000035 & 0.002957722 \\
\hphantom{0}1 & 0.9635960 & 0.002878444 \\
\hphantom{0}2 & 0.9557780 & 0.002895847 \\
\hphantom{0}3 & 0.9490003 & 0.002861895 \\
\hphantom{0}4 & 0.9448612 & 0.002879159 \\
\hphantom{0}5 & 0.9411233 & 0.002881512 \\
\hphantom{0}6 & 0.9377399 & 0.002861368 \\
[4pt]
\hphantom{0}7 & 0.9354613 & 0.002863227 \\
\hphantom{0}9 & 0.9345158 & 0.002864500 \\
10 & 0.9327541 & 0.002864051 \\
11 & 0.9319742 & 0.002867013 \\
12 & 0.9303069 & 0.002870322 \\
\hline
\end{tabular*}\vspace*{3pt}
\end{table}
%
\begin{figure}

\includegraphics{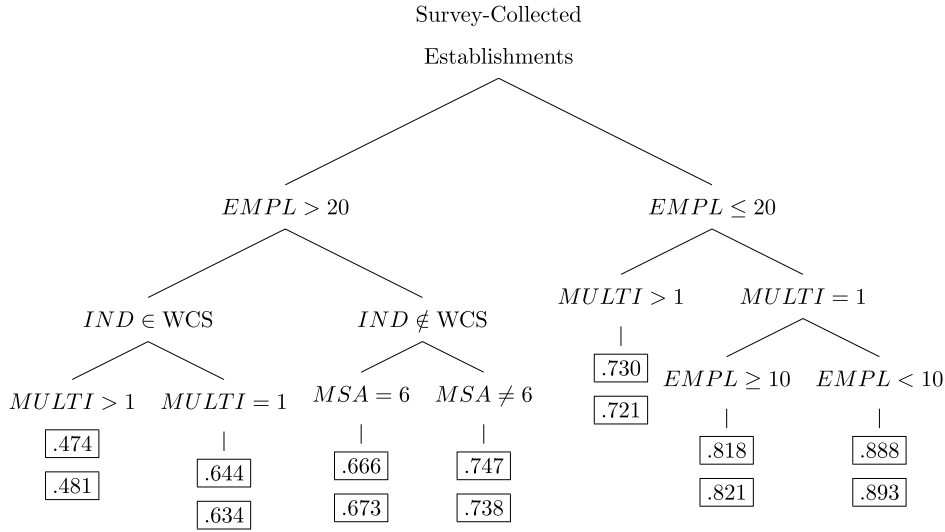}

\caption{This displays the regression tree estimating an establishment's
propensity to respond to the OES for a given set of characteristics.
This model was estimated using the procedure outlined in Section
\protect\ref{recursive partitioning} with May 2006 OES data.
The top value in the box is the estimated response rate for the May
2006 data used to build the regression tree model.
The bottom value is the realized response rate for the November 2006
OES sample, for each class of establishments
determined by the tree model using the May 2006 data.}
\label{response_tree}\vspace*{-1pt}
\end{figure}

Table~\ref{var_table_tree1} shows the estimated mean squared error of
the given tree by each successive split.
The relative prediction error is estimated using leave-out-$n/10$
cross-validation as described in Section~\ref{recursive partitioning}
[see \citet{HasTibFri01}; \citet{Sha93} for theory
behind cross-validation].
For example, the first row gives the mean of the 10 cross-validation
estimates for the
mean squared error divided by the mean squared error estimated from the
entire data set,
and the standard deviation of those 10 estimates for the tree with no splits.
The second row gives the same information for the tree with the one
split and so on.

The model was selected using the algorithm explained in Section~\ref{recursive partitioning}.
We can see from Table~\ref{var_table_tree1} that split 7 is the first
split for which
the absolute difference between its estimated mean squared error
(0.9354613) and the estimated mean squared error
of split 6 (0.9377399) is less than the estimator's estimated standard
error (0.00286).
Therefore, the resulting tree is the one comprised of the first six splits.
The above mean squared errors calculations are based on the unweighted
Bernoulli response model in equation (\ref{Bernoulli}).\vfill\eject

The resulting tree model is shown in Figure~\ref{response_tree}.
The tree gives the response rates for seven sets of establishments
defined by the splits on
establishment characteristics.
In the model we use the term \emph{white-collar service sector},
denoted $\mathbf{WCS},$ to
identify the establishments in the three industry super-sectors:
(1) Information, (2)~Finance, and (3) Professional and Business Services.
White-collar service sector industries as a group differed in response
rates to the OES compared to other industries.
This group was chosen automatically by the recursive partitioning
algorithm, as are all the splits in the model.

The model identifies the variables $\mathit{EMPL}$, $\mathit{IND}$, $\mathit{MULTI}$, and $\mathit{MSA}$
as having a significant impact on the propensity to respond for an
establishment.
Among small establishments, organizational complexity drives the
response rate.
Small, single unit establishments are most likely to respond to the
OES, in comparison to those that are part of
multi-unit firms.
In general, establishments with larger employment have lower response rates.
Specifically, for large establishments, the industry and the population
size of the metropolitan area
are important.
White-collar service establishments with a larger number of multi-units
have the lowest response rates.
In all other industries,
being located in a MSA with a population of one million or more is
associated with lower response rates.

\begin{figure}

\includegraphics{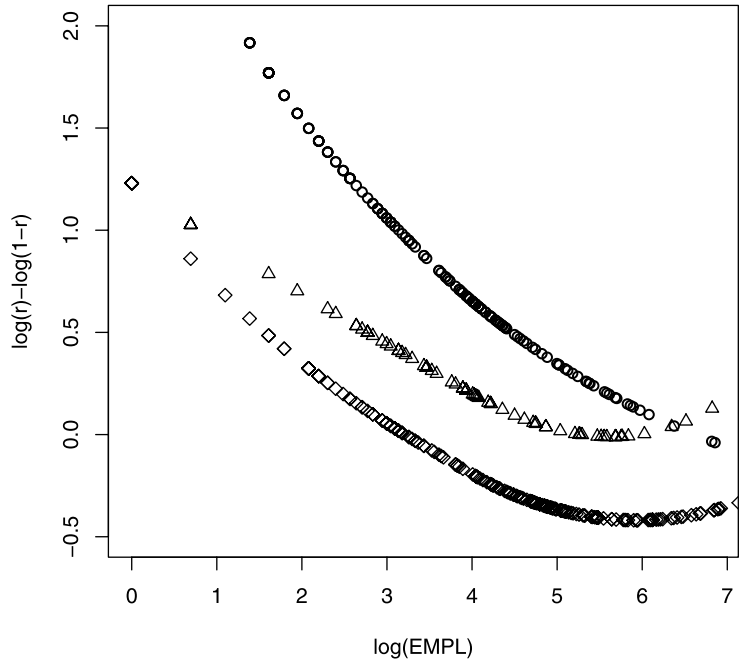}

\caption{The logit function $\operatorname{logit} (p(x) )=\log(p(x) )-\log
(1-p(x) )$ for the
smoothed response rates $r$ by the $\log$ transformed establishment size.
This is displayed for establishments in the professional and business
services industry category located in an MSA with over a million people.
The circles represent the log-odds ratio by log size for establishments
with $\mathit{MULTI}=1,$ the triangles are establishments with $\mathit{MULTI}=2,$ and
the diamonds are establishments with $\mathit{MULTI} \geq3.$
The response rate by transformed establishment size is estimated by a
loess smoother.}\label{smooth}
\end{figure}

In contrast, the logistic regression model,
%
\begin{eqnarray}\label{L1}
\operatorname{logit} (p(x) ) &= & \beta_0+ \beta_1 \log(\mathit{EMPL}) + \beta_2 \mathit{IND} +
\beta_3 \mathit{MSA} + \beta_4 \mathit{MULTI} \nonumber\\
&&{} + \beta_5 \log(\mathit{EMPL})*\mathit{IND} + \beta_6 \log(\mathit{EMPL})*\mathit{MSA} \nonumber
\\[-8pt]
\\[-8pt]
&&{} + \beta_7 \mathit{IND}*\mathit{MSA} + \beta_8 \mathit{IND}*\mathit{MULTI}\nonumber\\
&&{} + \beta_9 \log(\mathit{EMPL})*\mathit{IND}*\mathit{MSA},
\nonumber
\end{eqnarray}
is difficult to interpret.
Deciding on a logistic model in this situation, where there are a
number of continuous and categorical variables and the dependent variable
is associated with a number of interactions between the variables, is
nontrivial.
Even the best fitting logistic model [equation (\ref{L1})], obtained
using the stepwise model selection procedure,
does not seem to fit the data particularly well.
To see this, we consider establishments in the professional and
business services industry category located in an MSA with over a
million people.
Then, separately for establishments with $\mathit{MULTI}=1,$ $\mathit{MULTI}=2,$ and
$\mathit{MULTI}\geq3,$
we used a locally weighted smoother (LOESS) to fit the response rate
with respect to $\log(\mathit{EMPL})$
(see Figure~\ref{smooth}).
According to equation (\ref{L1}), the two curves representing
establishments with $\mathit{MULTI}=1$ and $\mathit{MULTI}=2$
should be linear with respect to $\log(\mathit{EMPL}).$
Looking at Figure~\ref{smooth}, the assumption of linearity seems
fairly plausible for establishments with $\mathit{MULTI}=1,$
but for establishments with $\mathit{MULTI}=2,$ the assumption seems invalid.
It should be noted that Figure~\ref{smooth} seems to imply that the
model would be improved by a quadratic term.
Any attempts to add this to the model, as well as additional attempts
at transforming variables,
led to a model that overfit the data.

To check the resulting tree model, we estimate the coefficients of its
simple function form twice,
once using the OES data from May 2006, and the second time using the
OES data from November 2006.
Because an establishment has probability zero of being selected for the
November survey if it was a sampled
unit in the May survey, the two data sets are mutually exclusive.
Estimated coefficients using the November data that are close to those
estimated from the May data would indicate that
the splits of the selected model accurately represent the effects that
certain establishment characteristics are likely to
have on an establishment's propensity to respond to future OES surveys.

%
\begin{table}
\tabcolsep=0pt
\caption{Results from the recursive partitioning of the OES mail survey data.
Column 1 displays coefficients for the set of splits estimating
response propensity for May 2006.
Column 2 displays November 2006 response coefficients, based on the May
tree model.
Column 3 displays coefficients for the tree model to estimate May 2006
average wage per employee}
\label{tab:simple_response}
\label{tab3}\begin{tabular*}{\textwidth}{@{\extracolsep{\fill}}ld{2.4}d{2.4}e{4.0}@{}}
\hline
& \multicolumn{1}{c}{\textbf{May response}} & \multicolumn
{1}{c}{\textbf{Nov. response}} & \multicolumn{1}{c@{}}{\textbf{May
wage}} \\
\textbf{Split} & \multicolumn{1}{c}{\textbf{coefficient}} &
\multicolumn{1}{c}{\textbf{coefficient}} & \multicolumn{1}{c@{}}{\textbf
{coefficient}} \\
\hline
$1$ & 0.8883 & 0.8933 & 8261 \\ 
$\mathit{EMPL} > 20$ & -0.1411 & -0.1556 & -970 \\ 
$\mathit{EMPL} > 20$  and  $\mathit{IND} \in\mathrm{WCS}$ & -0.1036 &
-0.1037 & 4818 \\ 
$\mathit{EMPL} > 20$  and  $\mathit{IND} \in\mathrm{WCS}$  and  $\mathit
{MULTI} > 1$ & -0.1691 & -0.1529 & 1298 \\ 
$\mathit{EMPL} > 20$  and  $\mathit{IND} \notin\mathrm{WCS}$  and  $\mathit
{MSA} = 6$ & -0.0810 & -0.0648 & 1706 \\ 
$\mathit{EMPL} \leq 20 $  and  $\mathit{MULTI} > 1$ & -0.1579 & -0.1722 &
3394 \\ 
$\mathit{EMPL} \leq 20 $  and  $\mathit{MULTI} = 1$  and  $
\mathit{EMPL} \geq 10$ & -0.0707 & -0.0720 & -559 \\ 
\hline
\end{tabular*}
\end{table}

The coefficients of the binary tree model given by equation (\ref{simp_eq})
are shown in Table~\ref{tab:simple_response}.
The first three columns of the table show the splits with the
corresponding coefficients estimating the propensity to respond
using the May and November data, respectively.
Comparing the two sets of estimated coefficients in this table, we see
the two estimates are quite close.
Indeed, comparing the estimated response rates to the rates obtained in
the November survey, shown in
Figure~\ref{response_tree},
we see that the model predicted within 1 percentage point of the
realized response rate for every group.
Since the November data set represents a completely disjoint set of
sampled units using the same
sample design as the data used to build the model, and considering the
very large sample size ($n=179\mbox{,}360$),
this test provides evidence that the regression tree model accurately describes
the association between the establishment characteristics and the
response propensity for the OES survey.
Note that, due to the large sample size in each panel of the survey,
all of the coefficients estimated are highly significant ($p$-value $\ll 0.001$).
Likewise, estimates for the standard errors of the coefficients are all
so small that\
they do not add much information and are therefore not reported.

\subsection{Centrally collected establishments}

Next, we applied the same procedure as above to analyze the response pattern
of establishments that were centrally collected.
We recursively partitioned the data of 8115 establishments included in
the May 2006 OES data
using the same set of characteristic variables as the survey-collected data.
The estimated mean squared error of the given tree by each successive
split is
summarized in Table~\ref{var_table_tree2}.

%
\begin{table}
\tabcolsep=0pt
\tablewidth=163pt
\caption{Relative estimated error by number of splits,
centrally collected establishments, May 2006}
\label{var_table_tree2}
\label{tab4}\begin{tabular*}{163pt}{@{\extracolsep{\fill}}lcc@{}}
\hline
\textbf{Split} & \textbf{Estimate} & \textbf{Standard error}\\
\hline
0 & 1.0002935 & 0.03239150 \\
1 & 0.9634402 & 0.03024989 \\
[4pt]
2 & 0.9441707 & 0.02935236 \\
3 & 0.9385307 & 0.02916412 \\
4 & 0.9263617 & 0.02840939 \\
5 & 0.9257025 & 0.02833990 \\
6 & 0.9258469 & 0.02834133 \\
\hline
\end{tabular*}
\end{table}
%
\begin{table}[b]
\tabcolsep=0pt
\caption{Results from the recursive partitioning of the OES centrally
collected survey data.
Column 1 displays coefficients for the set of splits estimating
response propensity for May 2006.
Column 2 displays November 2006 response coefficients, based on the May
tree model.
Column 3 displays coefficients for the tree model to estimate May 2006
average wage per employee}
\label{tab:simple_cc}
\label{tab5}\begin{tabular*}{\textwidth}{@{\extracolsep{\fill}}lccc@{}}
\hline
& \textbf{May response} & \textbf{Nov. response} & \textbf{May wage} \\
\textbf{Split} & \textbf{coefficient} & \textbf{coefficient} & \textbf
{coefficient} \\
\hline
$1$ & \hphantom{$-$}0.9590 & \hphantom{$-$}0.9538 & 8022 \\
$\mathit{MULTI} \leq 87$ & $-$0.1110 & $-$0.0855 & 1959\\
\hline
\end{tabular*}
\end{table}

\begin{figure}

\includegraphics{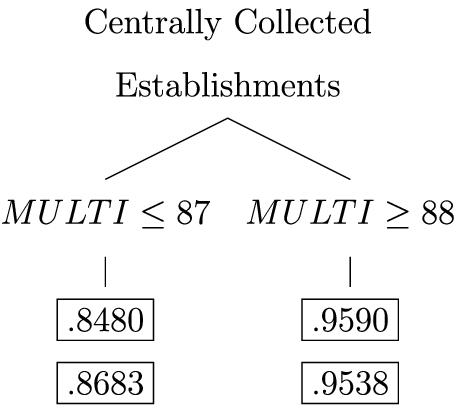}

\caption{Regression tree estimating an establishment's
propensity to respond to the OES for centrally collected units.
The top value in the box is the estimated response rate for the May
2006 data used to build the regression tree model.
The bottom value is the realized response rate for the November 2006
OES sample, for each class of establishments
determined by the tree model using the May 2006 data.}\label{response_cc_tree}
\end{figure}

The same model selection procedure resulted in the tree with one split
being selected.
The linear representation of the selected model is shown in Table~\ref{tab:simple_cc}.
The estimated coefficients using both May and November centrally
collected data are given in the first two columns.
Both models, the original using May data and the model using November
data to estimate coefficients,
are shown in Figure~\ref{response_cc_tree}.
Response rates are high for the central-collection mode.
Given the existing relationship that these firms have with BLS regional
offices to coordinate
with one single representative, it may be that centrally collected
firms have a more
comprehensive, centralized record systems.
In addition, respondent motivation in pursuing a central-collection
agreement, and the existing relationship
with an economist in the BLS regional office are likely to factor into
the high response rate.
Yet, comparing the coefficients, it is clear that the model is
consistently predicting a lower
response rate for establishments that are part of firms with a smaller
number of establishments.
This is in contrast with mail survey-collected establishments, where
establishments
that are part of more complex firms have a~lower response rate.

%

\section{Indication of wage bias}\label{bias}

One of the main objectives of the OES is to estimate wages for
different occupations and occupational groups.
When respondents to a survey differ in the outcomes being measured
compared to nonrespondents,
the survey results are likely to be biased.
In the last section, establishment characteristics were identified as
being strongly associated with
the propensity to respond to the OES.
In this section, we investigate the possibility that these
establishment characteristics
are also associated with wages.
If true, this would lead us to conclude that nonresponse is a~potential
source of bias in the OES wage estimates.

One difficulty in conducting nonresponse analyses is that outcome data
are unavailable for survey nonrespondents.
Therefore, we use 2005 second quarter administrative payroll data for
each establishment in the May 2006 OES sample
as provided to the BLS QCEW as a proxy.
Because the May 2006 establishment sample frame was derived from the
second quarter QCEW data in 2005,
these data provide the total number of employees and the total amount
of payroll wages paid
for every establishment selected into the May 2006 OES sample.
Since QCEW is a census, these administrative wage data are available
for both respondents and nonrespondents of the OES survey.
We consider the average wages paid per employee in the second quarter
for each establishment, by dividing
the reported total quarterly wages paid by the number of employees.
These data do not provide wages by occupation, nor account for the
number of hours worked as does the OES.
However, the reported amount should be associated with wages as
measured by OES, providing a good proxy.

\begin{figure}

\includegraphics{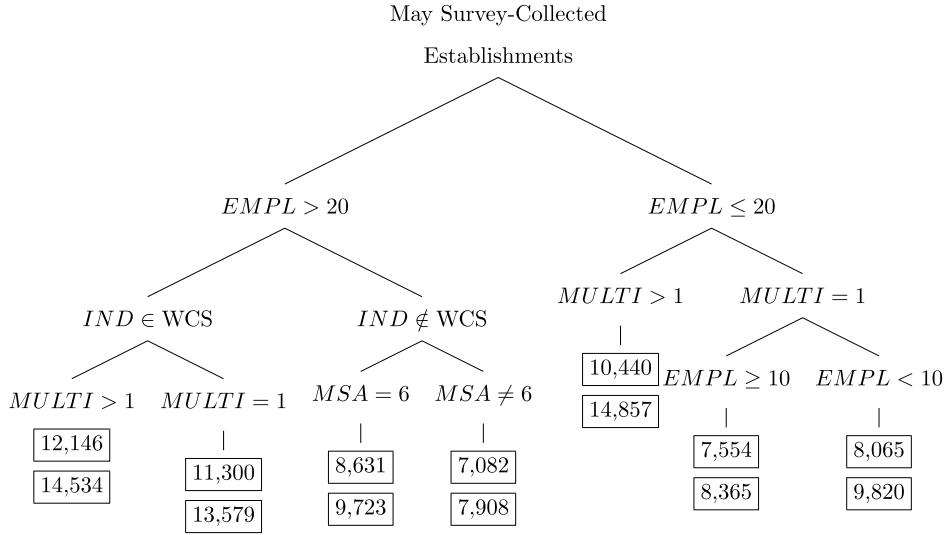}

\caption{Average quarterly wage per employee for the establishment
groups defined by the
regression tree model used to predict the response rates.
The top estimate is the average quarterly wage per employee at sampled
establishments that responded to
the OES survey, and the bottom estimate is for establishments that did
not respond to the OES survey.
The averages were calculated using the QCEW payroll data for the second
quarter of 2005.
This is the same administrative record data used as the frame to draw
the May 2006 sample.}
\label{wage_tree}
\end{figure}

Analysis of the wage data provides substantial evidence that
nonresponse could bias unadjusted wage estimates.
For example, the average wage paid per employee is \$8338 at
survey-collected establishments
that responded to the May 2006 OES, compared to an average of \$10,479
at establishments that did not respond.
The last column of Table~\ref{tab:simple_response} and Table~\ref{tab:simple_cc} gives the coefficients used to estimate the
average wage per employee for survey- and central-collection modes,
respectively.
These tables indicate that nonresponse (negative coefficients for the
response model) tends to coincide with higher pay
(positive coefficients for the wage model).
In addition, we fit the same regression tree models of establishment
characteristics used to model nonresponse
to the wage data for respondents and for nonrespondents separately.
The fitted models are shown in Figure~\ref{wage_tree}, with the average
wage per employee of
respondents in the top box and nonrespondents below.

The fitted model suggests that the interactions of establishment characteristics
associated with response propensity are also associated with the
average wage paid per employee.
Considering either the respondents or the nonrespondents separately,
the model tells a similar story.
Of the seven survey-collected establishment categories identified by
the tree model, the one with the lowest response rate,
large white-collar service establishments that are part of a
multi-establishment firm,
has an above average wage per employee.
The two categories with the highest response rates, establishments with
no more than 20 employees that
are not part of multi-establishment firms, have below average wages per
employee.

%

\begin{figure}

\includegraphics{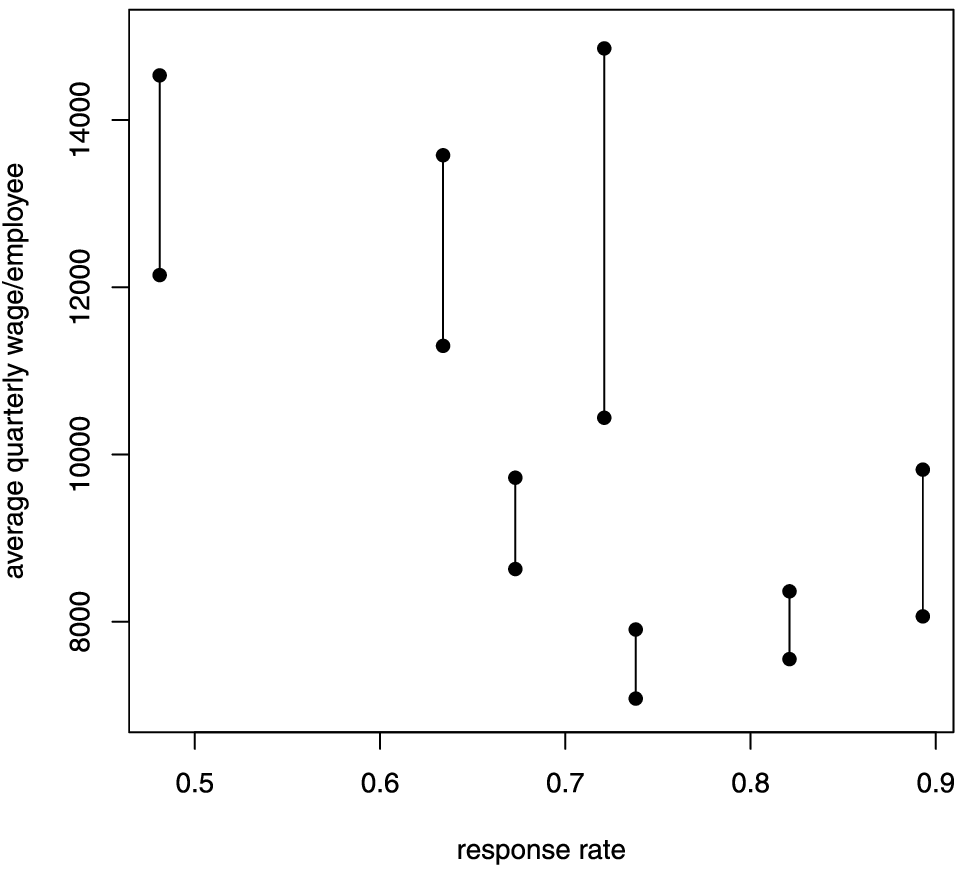}

\caption{For the seven categories of mail survey-collected
establishments defined
by the regression tree model, the average wage is plotted by response rate.
The average wage per employee is given for responding establishments
(bottom) and nonresponding establishments (top).
The line between the two estimates gives a visual representation of the
difference between responding and
nonresponding establishments within each category.
All wage estimates are for the second quarter of 2005 and are produced
from the QCEW records.}\label{wage_bias}
\end{figure}

Analyzing the differences in wages per employee between respondents and
nonrespondents within categories
suggests that there may be residual negative bias, even if the wage
estimates are adjusted.
If this difference persists for more refined models, the nonresponse
would be nonignorable.
Therefore, an effort to increase the response rate in certain
categories may be warranted.
For example, the model confirms that large, white-collar service
establishments are a potential
source of nonresponse bias.
The difference in the average wage between respondents and
nonrespondents in
this group is over \$2000, suggesting that more attention should be
addressed to these types of establishments.
However, large establishments outside of white-collar services may not
be as big a concern, despite their rather low response rates,
as the average wages of respondents and nonrespondents show much less
of a difference.
On the other hand, despite the modestly low response rate of
multi-establishments with
twenty employees or less, the large difference in the wage per employee
between respondents and nonrespondents
makes this a category deserving of more attention.
Figure~\ref{wage_bias} displays this difference in wages for
survey-collected respondents and nonrespondents for the seven
categories by response rate.
The difference is represented by a line between the two averages for
each category.
Three categories with below average response rates and relatively large
differences are evident in this graph:
large multi- and single-unit establishments in white-collar service
industries, and small multi-unit establishments.

\section{Discussion}\label{discussion}

Modeling establishment response rates using a regression tree model
allowed us to identify
important classes of establishments that have higher nonresponse rates
and pose a potential risk for nonresponse bias in the OES survey.
Unlike the groups formed from propensity score quantiles, these
interpretable groups are relevant for
testing theories on establishment nonresponse and forming future
adaptive data collection procedures.

Classes that pose the biggest risk have below average response rates
with relatively large differences in average wages per employee between
respondents and nonrespondents.
They include large (more than twenty employees) establishments in
white-collar service
industries and small (no more than twenty employees) establishments
that are part of a multi-establishment company.

\begin{figure}

\includegraphics{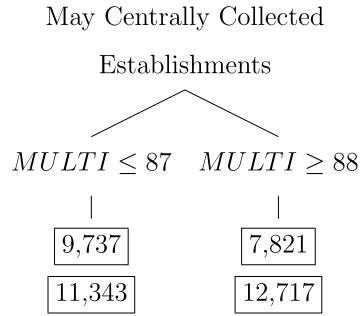}

\caption{Average quarterly wage per employee for centrally collected
establishments
using the payroll data for the second quarter of 2005.
The estimates are for the same tree model produced by the May 2006 OES
data for
estimating response rates of centrally collected establishments.
The top estimate is the average quarterly wage per employee at sampled
establishments that responded to
the OES survey, and the bottom estimate is for establishments that did
not respond to the OES survey.}\label{wage_cc_tree}
\end{figure}

Modeling the response rates for the two different modes of collection
shows that characteristics affecting establishment response are different
for the mail survey compared to the centrally collected establishments.
Given the higher response rate of centrally collected establishments,
arranging to have more
data collected using this method could increase the response rate.
This suggests a potential remedy for dealing with the risk posed by the
most problematic
category of establishments, those belonging to multi-establishment firms.
However, the large wage difference for respondents and nonresponders
in centrally collected establishments may limit the impact of this
solution on nonresponse bias (see Figure~\ref{wage_cc_tree}).
This is particularly the case for multi-unit firms with a larger number
of establishments.

%

%

%

The fact that the differences in average wage per employee between
respondents and nonresponders persist across
categories, even among centrally collected units, gives cause for
concern that the nonresponse bias could be nonignorable.
If so, adjusting for nonresponse using the administrative wage data as
well as the establishment characteristics
may help to reduce nonresponse bias.
Research on whether nonignorable nonresponse is a serious threat to the
OES wage data, as well as potential adjustment, is currently underway.

The study findings are strong and have many possible implications for
the OES survey program.
Nonresponse has distinct patterns in the OES, based on employment size,
industry,
multi-establishment status, and metropolitan location.
The OES program may want to consider survey design changes, such as
focused contact or nonresponse follow-up for
establishment groups with low response propensity and high wage differentials.
These types of changes could be integrated into a responsive design,
which OES is well set up to implement,
given its multiple mailings.
In addition, exploration of the QCEW average wage
as an auxiliary variable in nonresponse bias adjustments may be a
promising option.

OES may consider collecting more data via the central-collection mode
as a way to improve response rates in multi-establishment firms.
However, there are reasons for caution.
First, establishment respondents have made a special request to provide
data via a central collection arrangement.
Given the respondent self-selection involved, it is unclear whether
central collection could be implemented on a larger scale.
Second, because of the relatively large difference in wages for
responding and nonresponding centrally collected establishments,
changing to this mode of collection is likely to have a small impact on
bias reduction if this difference persists after the change.
Before attempting to expand this type of collection, a serious
assessment of how to reduce
nonresponse among large firms reporting by this mode would need to be
undertaken,
as well as a test of the viability of the data collection mode for
other multi-establishment firms.

\begin{appendix}
\section*{Appendix: Simulations to compare regression tree and logistic
regression modeling procedures}\label{sec:sim}

In this section we compare the performance of the regression tree modeling
to the more common logistic regression.
Specifically, we compare the nonparametric regression tree model
to a parametric model obtained using stepwise logistic regression in R.

To compare the two approaches,
we consider the accuracy of a given modeling procedure for predicting
an establishment's response propensity
when the response propensity $p(\mathbf{x})$ is a function of the given
values $\mathbf{x}.$
We test the two methods on five different functions for $p.$
For all five models we used randomly generated data containing six
independent variables
$\mathbf{x}=(x_1,x_2,x_3,x_4,c_1,c_2).$
Four variables, $x_1$, $x_2$, $x_3$, and $x_4,$
are integers uniformly distributed between 0 and 100 and two variables,
$c1$ and $c2$, are categorical variables.
The variable $c1$ has a binomial distribution with $n=4$ and $p= 0.2$ and
$c_2$ is Bernoulli with $p= 0.3.$
The random variable $\mathbf{R}_i=(R_{i1}, R_{i2}, R_{i3}, R_{i4},
R_{i5})$ was then generated as independent Bernoulli random
variables with $p=p(\mathbf{x})$ using the five models for $p(\mathbf
{x})$ described below.
Each generated data set had one hundred randomly generated points
$(\mathbf{R}_i, \mathbf{x}_i)$
using the above distribution.

The first of the five models for $p(\mathbf{x})$ used to compare
modeling procedures was
the simple logistic model with no interactions
\[
p_1(\mathbf{x})= \bigl(1+\exp\{- 0.003(1+3x_1-2x_2+3x_3)\} \bigr)^{-1}.
\]
The second model was also a logistic model with one quadratic term and
interactions among the variables.
The logistic model used was
\[
p_2(\mathbf{x})= \bigl(1+\exp\{- 0.0001(x_1+2x_2-3x_3-x_1x_2+2x_1x_3+x_3^2)\} \bigr)^{-1}.
\]
The third model
\[
p_3(\mathbf{x})= \bigl(1+\exp\{- 0.0001(c_1x_2x_3-x_1x_2+2x_1x_3+c_2x_3^2)\} \bigr)^{-1}
\]
is logistic with higher order interactions among the variables.
The fourth is the simple logistic model
\[
p_4(\mathbf{x})= \bigl(1+\exp\{- 0.01(x_1-x_2)\} \bigr)^{-1}
\]
for values of $\mathbf{x}$ with $x_1 > 35$ and with
\[
p_4(\mathbf{x})= \bigl(1+\exp\{- 0.01(2x_1+x_2)\} \bigr)^{-1}
\]
otherwise.
The fifth modeled $p(\mathbf{x})$ using the tree model\vspace*{3pt}
\begin{center}

\includegraphics{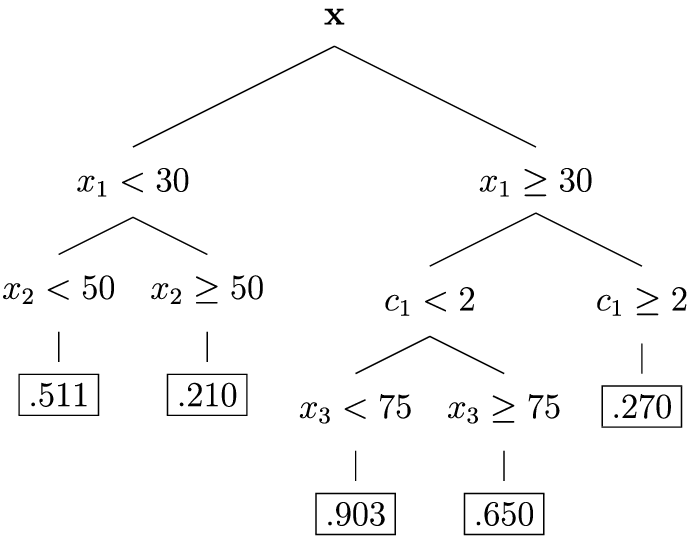}\vspace*{3pt}

\end{center}

\begin{figure}

\includegraphics{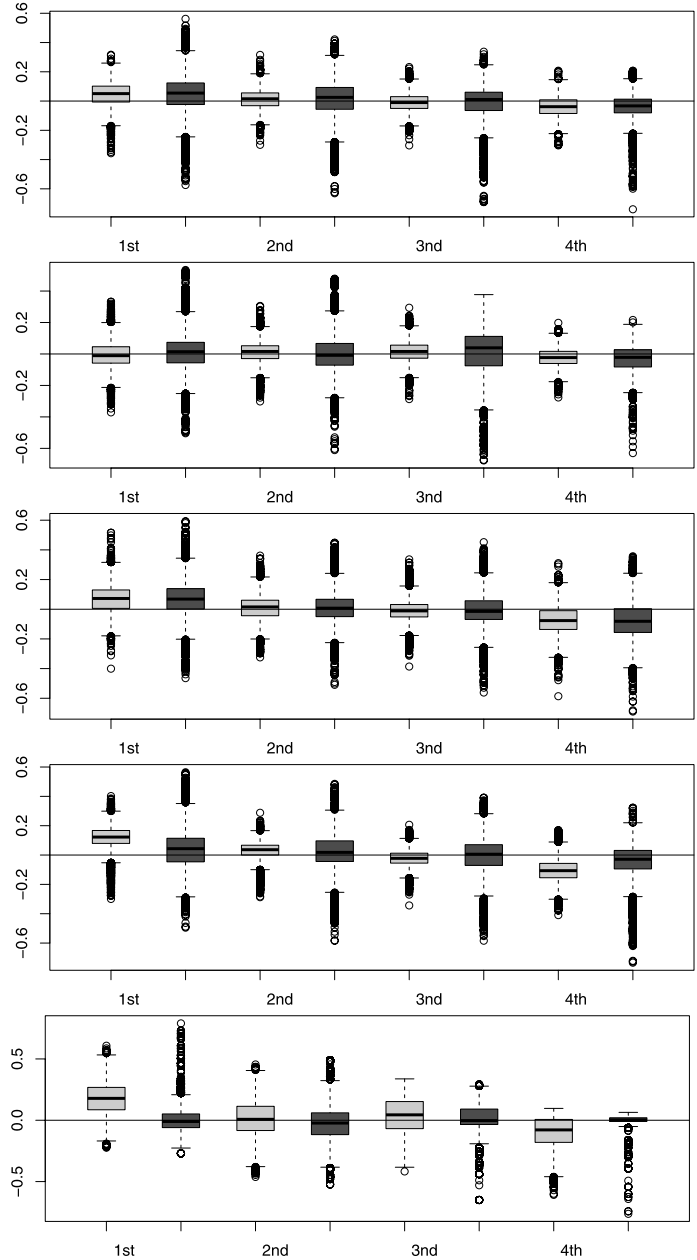}

\caption{Boxplots of the differences between the true value of
$p(\mathbf{x})$ and the predicted value for the log-linear model
(light gray) and the tree predictor (dark gray) for each of the four
quartiles of $p(\mathbf{x}).$
The whiskers are drawn at the most extreme data points that are within
1.5 times the inter-quartile range.
Outliers are defined as any point beyond the whiskers and are drawn on
the graph as circles.
The black horizontal line was drawn at zero to help see the skewness of
the errors for the different estimators.}
\label{fig:boxplots}
\end{figure}

For each simulated data set, we estimate the logistic model $\hat
{p}_{L}$ and the regression tree model $\hat{p}_T,$
for the response propensity for each of the five models of $p(x).$
A summary of the values of $p(\mathbf{x})$ produced by the five models
is given in Table~\ref{tab:models}.
Then for each model, we compare the predicted values~$\hat
{p}_{L}(\mathbf{x})$ and $\hat{p}_{T}(\mathbf{x})$ to the true model
$p(\mathbf{x}).$
For the five models, using 100 data sets of $n=500,$ boxplots of the
differences $\hat{p}_{L}(\mathbf{x})-p(\mathbf{x})$ and
$\hat{p}_{T}(\mathbf{x})-p(\mathbf{x})$ are given for each quantile of
$p(\mathbf{x})$ in Figure~\ref{fig:boxplots}.

%
\begin{table}[b]
\tabcolsep=0pt
\caption{Summary of values for $p(\mathbf{x})$, for the five models
used in the simulation}\label{tab:models}
\label{tab6}\begin{tabular*}{253pt}{@{\extracolsep{\fill}}lcccccc@{}}
\hline
\textbf{Model} & \textbf{Min.} & \textbf{1st Qu.} & \textbf{Median} &
\textbf{Mean} & \textbf{3rd Qu.} & \textbf{Max.} \\
\hline
1 & 0.36 & 0.58 & 0.65 & 0.64 & 0.71 & 0.86 \\
2 & 0.28 & 0.52 & 0.61 & 0.63 & 0.73 & 0.95 \\
3 & 0.27 & 0.48 & 0.53 & 0.55 & 0.61 & 0.97 \\
4 & 0.35 & 0.52 & 0.60 & 0.60 & 0.68 & 0.85 \\
5 & 0.21 & 0.51 & 0.65 & 0.65 & 0.90 & 0.90 \\
\hline
\end{tabular*}
\end{table}

%

The results show that both methods of modeling worked reasonably well
on all the data sets.
The logistic modeling performed slightly better when~$p(\mathbf{x})$
fit a log linear model but
worse than the tree model, as the $p(\mathbf{x})$ had discontinuities.
When using stepwise regression to find the best logistic model, we
searched over all models
using the six variables, one-way interactions, and quadratic terms.
Note that this was sufficient to fit model 1 and model~2 perfectly.
This would not be known in practice, and it is not clear how to choose
the number of interactions to include.
When too many interactions were included in the possible models, the
procedure performed less efficiently.
\end{appendix}


\section*{Acknowledgments}
The views expressed on statistical, methodological, technical, and operational
issues are those of the authors and not necessarily those of the U.S.
Bureau of Labor Statistics.
The authors thank the Editors and referees for their helpful comments,
as well as
John L. Eltinge and Jean Opsomer for comments and discussions that
improved this article.


%

\printaddresses


\begin{thebibliography}{23}

\bibitem[\protect\citeauthoryear{Breiman et al.}{1984}]{Breetal84}
%
\begin{bbook}[mr]
\bauthor{\bsnm{Breiman},~\bfnm{Leo}\binits{L.}},
\bauthor{\bsnm{Friedman},~\bfnm{Jerome~H.}\binits{J.~H.}},
\bauthor{\bsnm{Olshen},~\bfnm{Richard~A.}\binits{R.~A.}} \AND
\bauthor{\bsnm{Stone},~\bfnm{Charles~J.}\binits{C.~J.}}
(\byear{1984}).
\btitle{Classification and Regression Trees}.
\bpublisher{Wadsworth Advanced Books and Software}, \baddress{Belmont, CA}.
\bid{mr={0726392}}
\bptok{imsref}%
\end{bbook}
%
\endbibitem

\bibitem[\protect\citeauthoryear{Eltinge and Yansaneh}{1997}]{EltYan97}
%
\begin{barticle}[auto:STB|2012/01/18|07:48:53]
\bauthor{\bsnm{Eltinge},~\bfnm{J.}\binits{J.}} \AND
\bauthor{\bsnm{Yansaneh},~\bfnm{I.}\binits{I.}}
(\byear{1997}).
\btitle{Diagnostics for formation of nonresponse adjustment cells, with an
application to income nonresponse in the U.S. Consumer Expenditure Survey}.
\bjournal{Survey Methodology}
\bvolume{23}
\bpages{33--40}.
\bptok{imsref}%
\end{barticle}
%
\endbibitem

\bibitem[\protect\citeauthoryear{G\"{o}ksel, Judkins and Mosher}{1992}]{GokJudMos92}
%
\begin{barticle}[auto:STB|2012/01/18|07:48:53]
\bauthor{\bsnm{G{\"o}ksel},~\bfnm{H.}\binits{H.}},
\bauthor{\bsnm{Judkins},~\bfnm{D.}\binits{D.}} \AND
\bauthor{\bsnm{Mosher},~\bfnm{W.}\binits{W.}}
(\byear{1992}).
\btitle{Nonresponse adjustment for a telephone follow-up to a national
in-person survey}.
\bjournal{Journal of Official Statistics}
\bvolume{8}
\bpages{417--431}.
\bptok{imsref}%
\end{barticle}
%
\endbibitem

\bibitem[\protect\citeauthoryear{Gordon and Olshen}{1978}]{GorOls78}
%
\begin{barticle}[mr]
\bauthor{\bsnm{Gordon},~\bfnm{Louis}\binits{L.}} \AND
\bauthor{\bsnm{Olshen},~\bfnm{Richard~A.}\binits{R.~A.}}
(\byear{1978}).
\btitle{Asymptotically efficient solutions to the classification problem}.
\bjournal{Ann. Statist.}
\bvolume{6}
\bpages{515--533}.
\bid{issn={0090-5364}, mr={0468035}}
\bptok{imsref}%
\end{barticle}
%
\endbibitem

\bibitem[\protect\citeauthoryear{Gordon and Olshen}{1980}]{GorOls80}
%
\begin{barticle}[mr]
\bauthor{\bsnm{Gordon},~\bfnm{Louis}\binits{L.}} \AND
\bauthor{\bsnm{Olshen},~\bfnm{Richard~A.}\binits{R.~A.}}
(\byear{1980}).
\btitle{Consistent nonparametric regression from recursive partitioning
schemes}.
\bjournal{J. Multivariate Anal.}
\bvolume{10}
\bpages{611--627}.
\bid{doi={10.1016/0047-259X(80)90074-3}, issn={0047-259X}, mr={0599694}}
\bptok{imsref}%
\end{barticle}
%
\endbibitem

\bibitem[\protect\citeauthoryear{Hastie, Tibshirani and Friedman}{2001}]{HasTibFri01}
%
\begin{bbook}[mr]
\bauthor{\bsnm{Hastie},~\bfnm{Trevor}\binits{T.}},
\bauthor{\bsnm{Tibshirani},~\bfnm{Robert}\binits{R.}} \AND
\bauthor{\bsnm{Friedman},~\bfnm{Jerome}\binits{J.}}
(\byear{2001}).
\btitle{The Elements of Statistical Learning: Data Mining, Inference,
and Prediction}.
\bpublisher{Springer}, \baddress{New York}.
\bid{mr={1851606}}
\bptok{imsref}%
\end{bbook}
%
\endbibitem

\bibitem[\protect\citeauthoryear{Kim and Kim}{2007}]{KimKim07}
%
\begin{barticle}[mr]
\bauthor{\bsnm{Kim},~\bfnm{Jae~Kwang}\binits{J.~K.}} \AND
\bauthor{\bsnm{Kim},~\bfnm{Jay~J.}\binits{J.~J.}}
(\byear{2007}).
\btitle{Nonresponse weighting adjustment using estimated response probability}.
\bjournal{Canad. J. Statist.}
\bvolume{35}
\bpages{501--514}.
\bid{doi={10.1002/cjs.5550350403}, issn={0319-5724}, mr={2381396}}
\bptok{imsref}%
\end{barticle}
%
\endbibitem

\bibitem[\protect\citeauthoryear{Kott and Chang}{2010}]{KotCha10}
%
\begin{barticle}[mr]
\bauthor{\bsnm{Kott},~\bfnm{Phillip~S.}\binits{P.~S.}} \AND
\bauthor{\bsnm{Chang},~\bfnm{Ted}\binits{T.}}
(\byear{2010}).
\btitle{Using calibration weighting to adjust for nonignorable unit
nonresponse}.
\bjournal{J. Amer. Statist. Assoc.}
\bvolume{105}
\bpages{1265--1275}.
\bid{doi={10.1198/jasa.2010.tm09016}, issn={0162-1459}, mr={2752620}}
\bptok{imsref}%
\end{barticle}
%
\endbibitem

\bibitem[\protect\citeauthoryear{LeBlanc and Tibshirani}{1998}]{LeBTib98}
%
\begin{barticle}[auto:STB|2012/01/18|07:48:53]
\bauthor{\bsnm{LeBlanc},~\bfnm{M.}\binits{M.}} \AND
\bauthor{\bsnm{Tibshirani},~\bfnm{R.}\binits{R.}}
(\byear{1998}).
\btitle{Monotone shrinkage of trees}.
\bjournal{J. Comput. Graph. Statist.}
\bvolume{7}
\bpages{417--433}.
\bptok{imsref}%
\end{barticle}
%
\endbibitem

\bibitem[\protect\citeauthoryear{Little}{1982}]{Lit82}
%
\begin{barticle}[mr]
\bauthor{\bsnm{Little},~\bfnm{Roderick J.~A.}\binits{R.~J.~A.}}
(\byear{1982}).
\btitle{Models for nonresponse in sample surveys}.
\bjournal{J. Amer. Statist. Assoc.}
\bvolume{77}
\bpages{237--250}.
\bid{issn={0162-1459}, mr={0664675}}
\bptok{imsref}%
\end{barticle}
%
\endbibitem

\bibitem[\protect\citeauthoryear{Little}{1986}]{Lit86}
%
\begin{barticle}[auto:STB|2012/01/18|07:48:53]
\bauthor{\bsnm{Little},~\bfnm{R.}\binits{R.}}
(\byear{1986}).
\btitle{Survey nonresponse adjustments for estimates of means}.
\bjournal{International Statistical Review}
\bvolume{2}
\bpages{139--157}.
\bptok{imsref}%
\end{barticle}
%
\endbibitem

\bibitem[\protect\citeauthoryear{Little and Vartivarian}{2005}]{LitVar05}
%
\begin{barticle}[auto:STB|2012/01/18|07:48:53]
\bauthor{\bsnm{Little},~\bfnm{R.}\binits{R.}} \AND
\bauthor{\bsnm{Vartivarian},~\bfnm{S.}\binits{S.}}
(\byear{2005}).
\btitle{Does weighting for nonresponse increase the variance of survey means?}
\bjournal{Survey Methodology}
\bvolume{31}
\bpages{161--168}.
\bptok{imsref}%
\end{barticle}
%
\endbibitem

\bibitem[\protect\citeauthoryear{Opsomer and Miller}{2005}]{OpsMil05}
%
\begin{barticle}[mr]
\bauthor{\bsnm{Opsomer},~\bfnm{J.~D.}\binits{J.~D.}} \AND
\bauthor{\bsnm{Miller},~\bfnm{C.~P.}\binits{C.~P.}}
(\byear{2005}).
\btitle{Selecting the amount of smoothing in nonparametric regression
estimation for complex surveys}.
\bjournal{J. Nonparametr. Stat.}
\bvolume{17}
\bpages{593--611}.
\bid{doi={10.1080/10485250500054642}, issn={1048-5252}, mr={2141364}}
\bptok{imsref}%
\end{barticle}
%
\endbibitem

\bibitem[\protect\citeauthoryear{Petroni et al.}{2004}]{Petetal}
%
\begin{bmisc}[auto:STB|2012/01/18|07:48:53]
\bauthor{\bsnm{Petroni},~\bfnm{R.}\binits{R.}},
\bauthor{\bsnm{Sigman},~\bfnm{R.}\binits{R.}},
\bauthor{\bsnm{Willimack},~\bfnm{D.}\binits{D.}},
\bauthor{\bsnm{Cohen},~\bfnm{S.}\binits{S.}} \AND
\bauthor{\bsnm{Tucker},~\bfnm{C.}\binits{C.}}~%
(\byear{2004}).
\bhowpublished{Response rates and nonresponse in establishment
surveys--BLS and Census Bureau. In \textit{Federal Economic Statistics Advisory
Committee Meeting (December)}. Available at
\texttt{\href{http://www.bea.gov/about/pdf/ResponseratesnonresponseinestablishmentsurveysFESAC121404.pdf}%
{http://}
\href{http://www.bea.gov/about/pdf/ResponseratesnonresponseinestablishmentsurveysFESAC121404.pdf}%
{www.bea.gov/about/pdf/ResponseratesnonresponseinestablishmentsurveysFESAC121404.}
\href{http://www.bea.gov/about/pdf/ResponseratesnonresponseinestablishmentsurveysFESAC121404.pdf}%
{pdf}}.}
\bptok{imsref}%
\end{bmisc}
%
\endbibitem

\bibitem[\protect\citeauthoryear{Phipps and Jones}{2007}]{PhiJon07}
%
\begin{bmisc}[auto:STB|2012/01/18|07:48:53]
\bauthor{\bsnm{Phipps},~\bfnm{P.}\binits{P.}} \AND
\bauthor{\bsnm{Jones},~\bfnm{C.}\binits{C.}}
(\byear{2007}).
\bhowpublished{Factors affecting response to the occupational
employment statistics
survey.
In \textit{Proceedings of the 2007 Federal Committee on Statistical
Methodology Research
Conference}.
Available at \texttt{\href{http://www.fcsm.gov/07papers/Phipps.II-A.pdf}{http://www.fcsm.gov/07papers/}
\href{http://www.fcsm.gov/07papers/Phipps.II-A.pdf}{Phipps.II-A.pdf}}.}
\bptok{imsref}%
\end{bmisc}
%
\endbibitem

\bibitem[\protect\citeauthoryear{Rosenbaum and Rubin}{1983}]{RosRub83}
%
\begin{barticle}[mr]
\bauthor{\bsnm{Rosenbaum},~\bfnm{Paul~R.}\binits{P.~R.}} \AND
\bauthor{\bsnm{Rubin},~\bfnm{Donald~B.}\binits{D.~B.}}
(\byear{1983}).
\btitle{The central role of the propensity score in observational
studies for
causal effects}.
\bjournal{Biometrika}
\bvolume{70}
\bpages{41--55}.
\bid{doi={10.1093/biomet/70.1.41}, issn={0006-3444}, mr={0742974}}
\bptok{imsref}%
\end{barticle}
%
\endbibitem

\bibitem[\protect\citeauthoryear{Schouten and de Nooij}{2005}]{SchdeN}
%
\begin{bmisc}[auto:STB|2012/01/18|07:48:53]
\bauthor{\bsnm{Schouten},~\bfnm{B.}\binits{B.}} \AND\bauthor{\bparticle{de}
\bsnm{Nooij},~\bfnm{G.}\binits{G.}}
(\byear{2005}).
\bhowpublished{Nonresponse adjustment using classification trees.
Discussion Paper 05001, Statistics Netherlands. Available at
\texttt{
\href{http://www.cbs.nl/NR/rdonlyres/1245916E-80D5-40EB-B047-CC45E728B2A3/0/200501x10pub.pdf}%
{http://www.cbs.}
\href{http://www.cbs.nl/NR/rdonlyres/1245916E-80D5-40EB-B047-CC45E728B2A3/0/200501x10pub.pdf}%
{nl/NR/rdonlyres/1245916E-80D5-40EB-B047-CC45E728B2A3/0/200501x10pub.pdf}}}.
\bptok{imsref}%
\end{bmisc}
%
\endbibitem

\bibitem[\protect\citeauthoryear{Shao}{1993}]{Sha93}
%
\begin{barticle}[mr]
\bauthor{\bsnm{Shao},~\bfnm{Jun}\binits{J.}}
(\byear{1993}).
\btitle{Linear model selection by cross-validation}.
\bjournal{J. Amer. Statist. Assoc.}
\bvolume{88}
\bpages{486--494}.
\bid{issn={0162-1459}, mr={1224373}}
\bptok{imsref}%
\end{barticle}
%
\endbibitem

\bibitem[\protect\citeauthoryear{Tomaskovic-Devey, Leiter and Thompson}{1994}]{aa}
%
\begin{barticle}[auto:STB|2012/01/18|07:48:53]
\bauthor{\bsnm{Tomaskovic-Devey},~\bfnm{D.}\binits{D.}},
\bauthor{\bsnm{Leiter},~\bfnm{J.}\binits{J.}} \AND
\bauthor{\bsnm{Thompson},~\bfnm{S.}\binits{S.}}
(\byear{1994}).
\btitle{Organizational survey nonresponse}.
\bjournal{Administrative Science Quarterly}
\bvolume{39}
\bpages{439--457}.
\bptok{imsref}%
\end{barticle}
%
\endbibitem

\bibitem[\protect\citeauthoryear{Toth and Eltinge}{2011}]{TotEltN1}
%
\begin{barticle}[auto:STB|2012/01/18|07:48:53]
\bauthor{\bsnm{Toth},~\bfnm{D.}\binits{D.}} \AND
\bauthor{\bsnm{Eltinge},~\bfnm{J.}\binits{J.}}
(\byear{2011}).
\btitle{Building consistent regression trees from complex
sample data}.
\bjournal{J. Amer. Statist. Assoc.}
\bvolume{106}
\bpages{1626--1636}.
\bptok{imsref}%
\end{barticle}
%
\endbibitem

\bibitem[\protect\citeauthoryear{Toth and Eltinge}{2008}]{TotEltN2}
%
\begin{bmisc}[auto:STB|2012/01/18|07:48:53]
\bauthor{\bsnm{Toth},~\bfnm{D.}\binits{D.}} \AND
\bauthor{\bsnm{Eltinge},~\bfnm{J.}\binits{J.}}
(\byear{2008}).
\bhowpublished{Simple function representation of regression trees.
Bureau of Labor Statistics Technical Report}.
\bptok{imsref}%
\end{bmisc}
%
\endbibitem

\bibitem[\protect\citeauthoryear{Vartivarian and Little}{2002}]{VarLit}
%
\begin{bmisc}[auto:STB|2012/01/18|07:48:53]
\bauthor{\bsnm{Vartivarian},~\bfnm{S.}\binits{S.}} \AND
\bauthor{\bsnm{Little},~\bfnm{R.}\binits{R.}}
(\byear{2002}).
\bhowpublished{On the formation of weighting adjustment cells for unit
nonresponse. In
\textit{Proceedings of the Survey Research Methods Section}
3553--3558. Amer. Statist. Assoc., Alexandria, VA}.
\bptok{imsref}%
\end{bmisc}
%
\endbibitem

\end{thebibliography}
\end{document}